\documentclass{article}

\usepackage[utf8]{inputenc} 
\usepackage[T1]{fontenc}    
\usepackage{hyperref}       
\usepackage{url}            
\usepackage{booktabs}       
\usepackage{amsmath,amsfonts,amssymb}       
\usepackage{nicefrac}       
\usepackage{microtype}      
\usepackage{lipsum}
\usepackage{graphicx}
\usepackage{authblk}

\title{Formation of jumps in isentropic flows. Classification of solitary shock waves.}
\author[1]{Kirill Karelsky}
\author[1,2]{Arakel Petrosyan\thanks{apetrosy@iki.rssi.ru}}

\affil[1]{Space Research Institute of the Russian Academy of Sciences, Moscow, Russia}
\affil[2]{Moscow Institute of Physics and Technology, Moscow, Russia}

\begin{document}
\maketitle

\begin{abstract}
The conditions for an arbitrary jump occurrence in isentropic flow are studied. It is shown that the jump in gas-dynamic parameters arises as a result of the evolution of a self-similar flow. The concept of self-focusing Riemann waves is introduced. It is shown that an arbitrary jump is formed only by these waves and the conditions for its generation are found. It is shown that there exists a critical velocity, below which a discontinuity cannot be formed isentropically. Also found is the second critical value of velocity, exceeding which a discontinuity is formed only in the presence of a vacuum region. It is shown that there are only two classes of solitary shock waves: those that form in a medium containing a vacuum region and those that form in a continuous medium. It is shown that not every fall of the Riemann wave leads to the appearance of a shock wave.
\end{abstract}


\section{Introduction}
Appearance of jumps in gas dynamics is a key problem for understanding natural processes and their mathematical modeling. The concept of an arbitrary discontinuity is widely used in gas dynamics and is key, for example, in the formulation of the Riemann fundamental problem of discontinuity decay \cite{riemann1860fortpflanzung}. An arbitrary discontinuity in gas dynamics is considered as given, leaving open the question of the possibility of its formation by an isentropic process. It is shown that there exists a critical velocity, below which a discontinuity cannot be formed isentropically. Also found is the second critical value of velocity, exceeding which a discontinuity is formed only in the presence of a vacuum region.  Of particular interest are solitary shock fronts, as physical realizations of stable jumps in a continuous medium. The main known results are obtained for the propagation and interaction of solitary shock fronts with other perturbations, leaving open the question of their origin.

Let us discuss the fundamental problem of the generation of a solitary shock front. By a solitary shock front we imply a special case of a stable jump in physical characteristics of the flow of a continuous medium. We consider a solitary shock front together with a contact discontinuity. The contact discontinuity and the shock front are integral parts of the shock wave as a physical phenomenon. It is this representation that makes it possible to identify the observed physical process as a shock wave. Indeed, in a solitary shock wave, the contact discontinuity always accompanies the shock front in an isentropic continuous medium. Behind the shock front, the entropy of the flow increases with respect to the initial isentropic flow, and the contact discontinuity ensures that entropy is constant outside the region of the shock wave. An important consequence of this consideration of a shock wave is the existence of a single point of its origin, which simultaneously belongs to the shock front and the contact discontinuity, that is, the points of their intersection. The presence of such an intersection point, together with the constancy of the flow at infinity (the condition for shock wave compactness) ensures the self-similarity of the flow characteristics near it. Thus, the analysis of self-similar solutions near such points allows one to investigate the transformation of isentropic flows into a solitary shock wave and propose a classification of shock waves based on their individual physical properties. 

Note that the inherent connection of the shock front with the contact discontinuity in isentropic gas-dynamic flows is a consequence of non-trivial thermodynamics in polytropic media. Nevertheless, in some cases of quasilinear hyperbolic systems that exclude thermodynamic effects from the description, a shock front exists without contact discontinuity. A typical example of such systems is the classical shallow water equations, in which the shock front is called the hydrodynamic jump.
Our study is based on the analysis of partial self-similar solutions for equations of gas dynamics of a polytropic gas and the solution of an arbitrary discontinuity decay problem (the Riemann problem). The paper consists of two parts. First, we find the conditions for the occurrence of discontinuities in isentropic flows and the structure of the currents that form this discontinuity. Then we find the condition under which an arbitrary discontinuity evolves into a solitary shock wave. Using the conditions for the formation of discontinuities together with the condition for the transformation of a discontinuity into a shock wave, we find a classification of shock waves based on the structure of the flows generating them.

In this paper we show: 1) all possible types of perturbations in isentropic processes forming jumps; 2) conditions for the realization of each of the perturbations obtained; 3) that any solitary shock wave can be formed by described perturbations; 4) the criterion for splitting shock waves into two classes, determined solely by their own physical properties, intensity and environment rheology; 5) a falling Riemann wave is presented that does not violate the flow isentropy.

\section{The jump in gas-dynamic parameters in a continuous medium as a result of the self-similar flows evolution. }

\begin{figure}
    \centering
    \includegraphics[width=.7\linewidth]{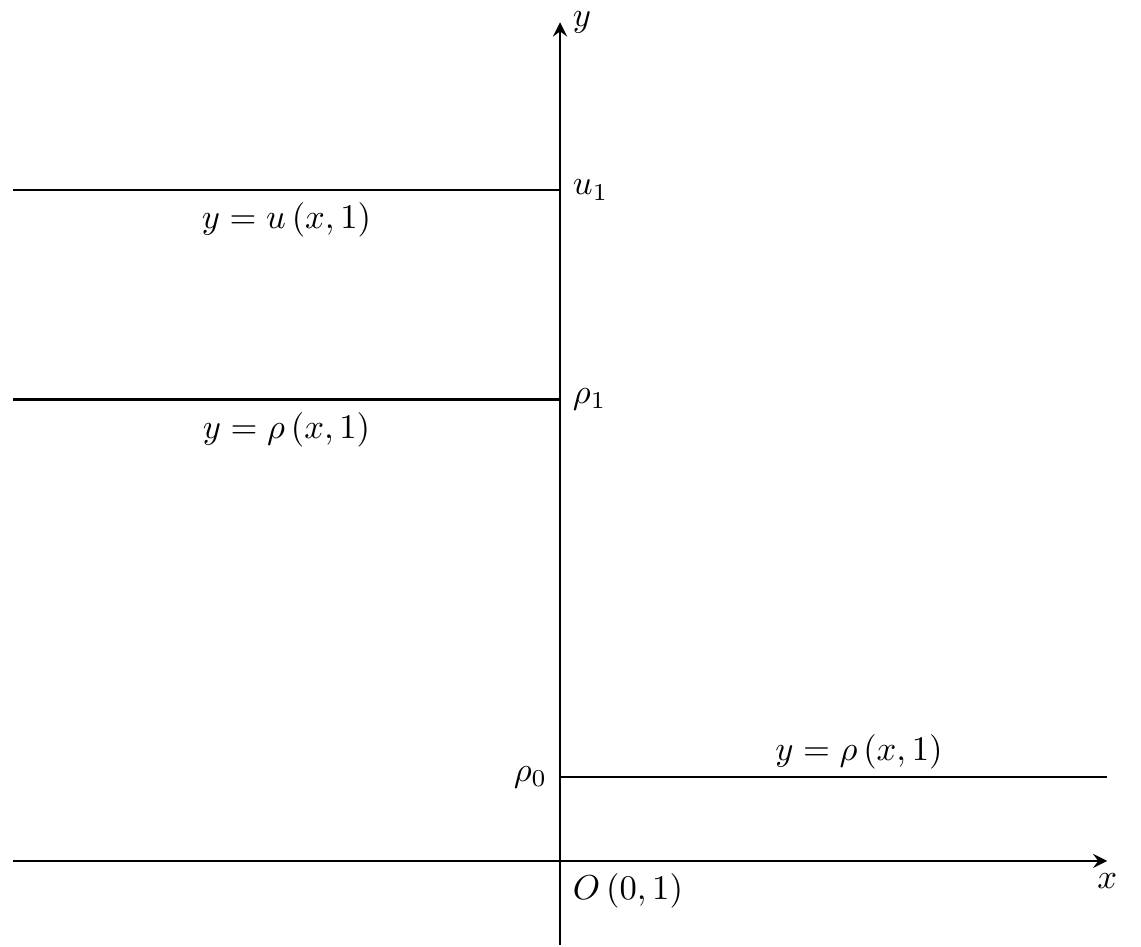}
    \caption{Arbitrary jump in the selected coordinate system at $t=1, \hat{u}(x,1)=\hat{u}_1, \hat{\rho}(x,1)=\hat{\rho}_1$ for $x<0$ and $\hat{u}(x,1)=\hat{u}_0, \hat{\rho}(x,1)=\hat{\rho}_0$ for $x \geq 0$.}
    \label{fig:fig1}
\end{figure}

\begin{figure}
    \centering
    \includegraphics[width=.7\linewidth]{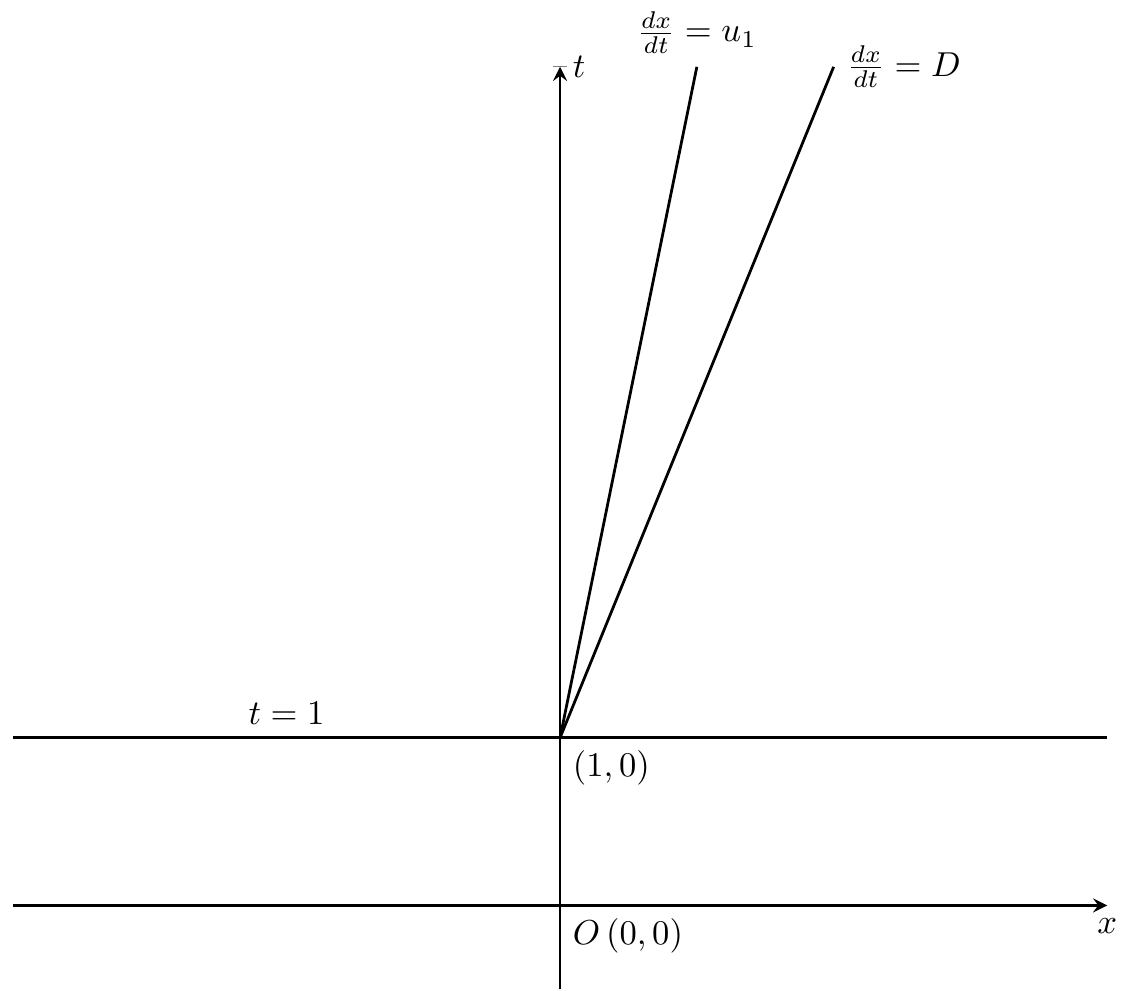}
    \caption{A solitary shock wave propagating to the right through a resting gas. $u_1>0$ – contact discontinuity propagation velocity, $D > 0$ – shock wave front propagation velocity.}
    \label{fig:fig2}
\end{figure}

To find the structure of the flows that form an arbitrary jump, we find all possible particular self-similar solutions converging to a point: shock waves, contact discontinuities, and two self-focusing waves facing forward and backward. The isentropy of the flow and the homogeneity of the medium exclude contact discontinuities and shock waves. The remaining self-focusing waves form the following wave patterns: only left or only right self-focusing waves, two self-focusing waves (left and right) separated by either a vacuum zone or a constant flow zone. We show that it is the last two pictures that form solitary shock waves. The conditions for their implementation are determined by the consistency of the characteristic picture, specifically the characteristics of particular solutions must pass in the correct sequence. Flows that form an arbitrary discontinuity are also obtained from solving an arbitrary discontinuity decay problem by replacing the time variable with a minus time in any arbitrarily small neighborhood of the singularity point at which the jump is formed. 
In this section all elementary self-similar perturbations converging to a given point are considered and all possible configurations of their combinations in isentropic processes are studied.
Let us consider a solitary shock wave as a special case of an arbitrary non-zero jump in the density of a homogeneous continuous medium $S=S_0$:
\begin{equation*}
    \left(\hat{u}_{a},\hat{\rho}_{a}\right),\left(\hat{u}_{b},\hat{\rho}_{b}\right),
\end{equation*}
where $\hat{u}_{a},\hat{u}_{b}$ are velocities and $\hat{\rho}_{a},\hat{\rho}_{b}$ are densities of continuous medium at the jump point and $\hat{\rho}_{a} \ne \hat{\rho}_{b}$.
The coordinate system is set up so that so that the region containing lower density medium is on the right and the origin $O\left( 0,0 \right)$ is chosen so that the medium in this region is at rest (Fig. \ref{fig:fig1}). In our notation we indicate parameters with index 1 to the left of discontinuity, and index 0 — to the right.
\begin{equation*}
    \left(\hat{u}_{1},\hat{\rho}_{1}\right),\left(\hat{u}_{0},\hat{\rho}_{0}\right),
\end{equation*}
where $\hat{\rho}_{1} > \hat{\rho}_{0}$ and $\hat{u}_{0}=0$.
A special case of such a jump is a solitary shock wave. According to the second law of thermodynamics, the shock front must run on a stationary medium, since shock waves of expansion cannot exist, and the medium density on the left is greater than on the right. Since the origin of a shock wave is considered in a homentropic medium, and the entropy at the front must increase, the region behind the shock front must be separated from the region of constant entropy. Thus, the front of a solitary shock wave is followed by its integral part - the contact gap. The flow configuration thus obtained is a solitary shock wave (Fig. \ref{fig:fig2}), where $u_1$ is a velocity of contact discontinuity and $\rho_{1}=\hat{\rho}_1$, $\rho_{0}=\hat{\rho}_0$ are densities of continuous medium.  We note that, by the definition of a contact gap, its velocity decreases with both the velocity inside the shock wave and the velocity of the unperturbed flow on the left. Thus, in the chosen coordinate system, an arbitrary density discontinuity becomes a shock wave, when $\hat{u}_{1}=u_1$.  Since the shock front moves in a positive direction along a stationary medium, the contact gap velocity is strictly positive. We determine the value of $u_1$  from the Riemann problem solution \cite{riemann1860fortpflanzung,rozhdestvenski_1983systems}. The solution contains five different configurations, with only two configurations containing shock waves for a dense flow in the positive direction: 1) an expansion wave and a right shock wave, 2) two shock waves. With the velocity of the incident flow increase, the intensity of the expansion wave in the first configuration lowers up to its complete degeneration. The complete degeneration of the expansion wave can be formally treated as the appearance of a left shock wave of zero intensity. (Further increase in the incident flow speed entails an increase in the intensity and transition to the configuration 2). The boundary value of the incident flow velocity separating these configurations determines the right solitary wave \cite{courant1948}:
\begin{equation}  \label{eqn:ineq1}
    u_1=\left( 1-\varkappa \right) c_0 \left( \sqrt{\frac{ \left(\frac{\rho_1}{\rho_0}\right)^\gamma +\varkappa }{1+\varkappa}} - \sqrt{\frac{1+\varkappa}{ \left(\frac{\rho_1}{\rho_0}\right)^\gamma +\varkappa }} \right),
\end{equation}
where $\varkappa=\frac{\gamma-1}{\gamma+1}$, $\gamma$ is adiabatic index and $c_0$ is speed of sound ahead of the shock wave front.

These arguments reduce the problem of a solitary shock wave formation to finding of an isentropic flow perturbation that converges to a point and connects two semi-infinite regions of constant homentropic flows. Therefore, to solve the problem, it is necessary to use flows with characteristics converging to a point and adjacent to constant flows. Characteristics of constant flows are parallel straight lines, and hence the desired perturbation should have straight line characteristics as well. There are only three types of flows in polytropic media that have such characteristics, namely shock waves, contact gaps and Riemann waves. Shock waves and contact gaps cannot be used due to the requirements of homentropy and homogeneity. Thus, the formation of an arbitrary density jump and a solitary shock wave in particular is determined by a perturbation consisting exclusively of Riemann waves converging to a point. The requirement of characteristics convergence limits the class of possible perturbations leaving only compression waves. The condition of convergence of Riemann waves to a single point is equivalent to their self-similarity. We call self-similar compression waves by pseudo-centered Riemann waves by analogy with the well-known centered Riemann waves emanating from one point and being expansion waves.

Pseudo-centered Riemann waves can be obtained from centered Riemann waves by a formal change of time $t$  by the opposite value $-t$. Evidently, pseudo-centered waves inherit all the properties of centered Riemann waves to within this substitution. There are two types of pseudo-centered waves: pointed forward or propagating along characteristics $\frac{dx}{dt}=u+c$  and pointed backwards, propagating along $\frac{dx}{dt}=u-c$.  Forward waves preserve the Riemann invariant $R=u-\frac{2}{\gamma-1}c$ and backward waves preserve invariant $S=u+\frac{2}{\gamma-1}c$  in the region of propagation. The velocity in pseudo-centered waves decreases linearly regardless of the propagation direction, and the speed of sound retaining a linear dependence grows in the backward wave and falls in the forward wave. Just as in the case of centered waves, no more than one pseudo-centered wave of one type can converge at one point. Due to the obvious inequality $u+c>u-c$, the sequence of waves of different classes is rigidly regulated - a backward wave follows a forward wave. Thus, there are four types of perturbations converging at one point and bordering on constant flows: 1) a forward pseudo-centered Riemann wave, 2) a backward pseudo-centered Riemann wave, 3) a configuration of two pseudo-centered Riemann waves - forward and backward, and 4) the configuration of two pseudo-centered Riemann waves separated by vacuum. The first two types of perturbations cannot be implemented to form a solitary shock wave. The first one is out due to the well-known relations between polytropic and Hugoniot shock adiabats, and the second because of the growth of sound speed and with it the density inside the wave. Let us obtain the conditions under which each of the latter two configurations that form a solitary shock wave is realized. Since the solitary shock wave is a special case of an arbitrary discontinuity, we first determine the conditions for the formation of an arbitrary discontinuity. Indeed, the four types of perturbations obtained are the only possible ones in isentropic processes that form a jump in polytropic media. Violation of the jump formation isentropy extends the class of perturbations converging at one point and bordering on constant currents to an infinite set, due to the fronts of shock waves pointed backward and forward. The number of such fronts converging into a point is not limited, except their sequence is regulated: any front of the left shock wave is located to the right of any front of the left shock wave. In addition, the presence of a shock front makes a joint configuration of disturbances containing two or more self-focusing waves of the same class. Thus, in the general case of a non-isentropic process, a perturbation consists of a set of shock fronts, a contact discontinuity, and self-focusing waves of both classes converging at a point.

Returning to isentropic processes, it is important to note that an arbitrary jump can also be formed by the first two types of disturbances corresponding to single self-focusing waves. However, the first two cases do not need to be considered separately, since for the formation of an arbitrary discontinuity they are a special case of the third perturbation, the merging of R and S self-focusing waves. In addition, the strict inequality in the density ratio at an arbitrary jump is redundant and is used only in the structure of a solitary shock wave formation.

Consider point $\left(0,t_0\right)$  as an origin of the jump. Denote the rays corresponding to the extreme characteristics of self-focusing waves, as $OA, OB, OC, OD$. These rays break the lower half-plane into five areas, which are numbered in Roman numerals, from left to right.

Thus, all possible configurations of self-similar perturbations are obtained, which form a jump at a given point $O\left(0,t_0\right)$  as a result of the isentropic process.

\section{Generation of a jump by two self-focusing Riemann waves, separated by a vacuum region.}
In this section, we obtain the realization conditions of a perturbation configuration consisting of two self-focusing waves and a vacuum zone between them.

\begin{figure}
    \centering
    \includegraphics[width=.7\linewidth]{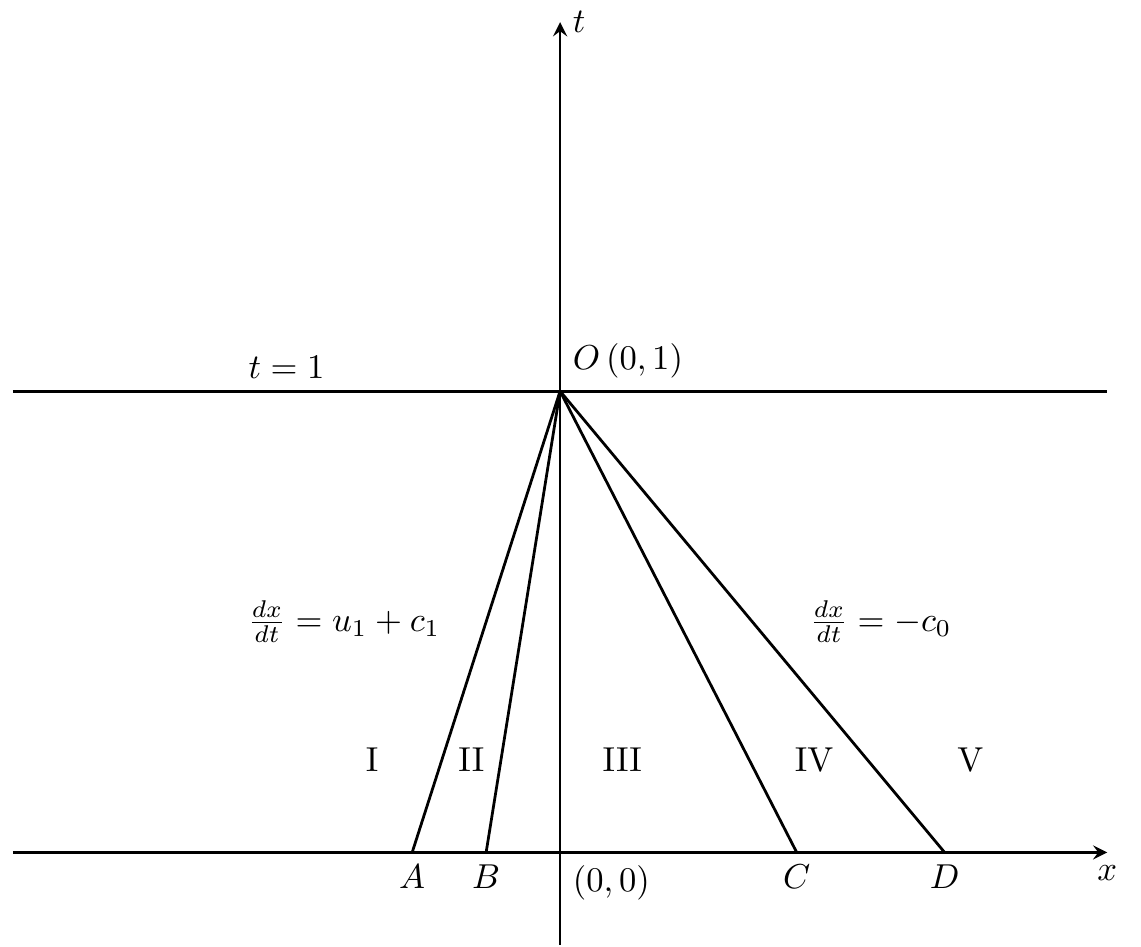}
    \caption{Cumulation of two self-focusing Riemann waves separated by a vacuum region III, $\rho \equiv 0$.}
    \label{fig:fig3}
\end{figure}

Let us indicate the values of physical quantities in each obtained region (Fig. \ref{fig:fig3}). In region I, we have a constant flow and the following variable values: $u=u_1,p=p_1,c=c_1$. In region II, there is a self-focusing Riemann R-wave and, therefore, the following relations hold:
\begin{equation*}
    R \equiv u-\frac{2}{\gamma-1}c=\text{const}
\end{equation*}
\begin{equation*}
    u_{II}-\frac{2}{\gamma-1}c_{II}=u_{1}-\frac{2}{\gamma-1}c_{1}
\end{equation*}
Region III is vacuum and therefore $p_{III} \equiv 0, c_{III} \equiv 0$.
In region IV there is a self-focusing Riemann S-wave and the following relations hold:
\begin{equation*}
    S \equiv u+\frac{2}{\gamma-1}c=\text{const}
\end{equation*}
\begin{equation*}
    u_{IV}+\frac{2}{\gamma-1}c_{IV}=u_0+\frac{2}{\gamma-1}c_0=\frac{2}{\gamma-1}c_0
\end{equation*}
Region V is a constant flow: $u=u_0=0,p=p_0,c=c_0$.
Equations defining rays $OA, OB, OC, OD$ have the form:
\begin{equation*}
    \frac{dx}{dt}=u_1+c_1,\quad \frac{dx}{dt}=u_1-\frac{2}{\gamma-1}c_1,\quad \frac{dx}{dt}=\frac{2}{\gamma-1}c_1,\quad \frac{dx}{dt}=c_0
\end{equation*}
Let us write the conditions for this configuration compatibility:
\begin{equation*}
    u_1+c_1>u_1-\frac{2}{\gamma-1}c_1>\frac{2}{\gamma-1}c_0>c_0
\end{equation*}
Considering $c_1 \geq c_0$ and $u_1>0$  the only non-trivial condition is $u_1-\frac{2}{\gamma-1}c_1>\frac{2}{\gamma-1}c_0$. Thus $u_1>\frac{2}{\gamma-1}\left( c_0+c_1 \right)$ and finally:
\begin{equation} \label{eqn:ineq2}
    u_1>\frac{2}{\gamma-1}c_1\left( 1+c' \right)
\end{equation}
where $c'=c_1 /c_0 $ is ratio of velocities of weak perturbations at the jump.

Given the thermodynamic relations provided by the condition of polytropic medium:
\begin{equation*}
    p'={c'}^{\frac{2\gamma}{\gamma-1}}, \quad c'={p'}^{\frac{\gamma-1}{2\gamma}}
\end{equation*}
where $p'=p_1 / p_0$ is ratio of pressure at the jump, inequality 2 takes the form:
\begin{equation}  \label{eqn:ineq3}
    u_1>\frac{2}{\gamma-1}c_1\left( 1+{p'}^{\frac{1-\gamma}{2\gamma}} \right)
\end{equation}
Thus, at velocities satisfying the found inequality \ref{eqn:ineq3}, all the jumps are formed by a flow containing a vacuum region.

\section{Generation of a jump by two self-focusing Riemann waves in a simply connected region.}

In this section, we obtain the realization conditions of a perturbations configuration consisting of two self-focusing waves separated by a zone of constant flow.

\begin{figure}
    \centering
    \includegraphics[width=.7\linewidth]{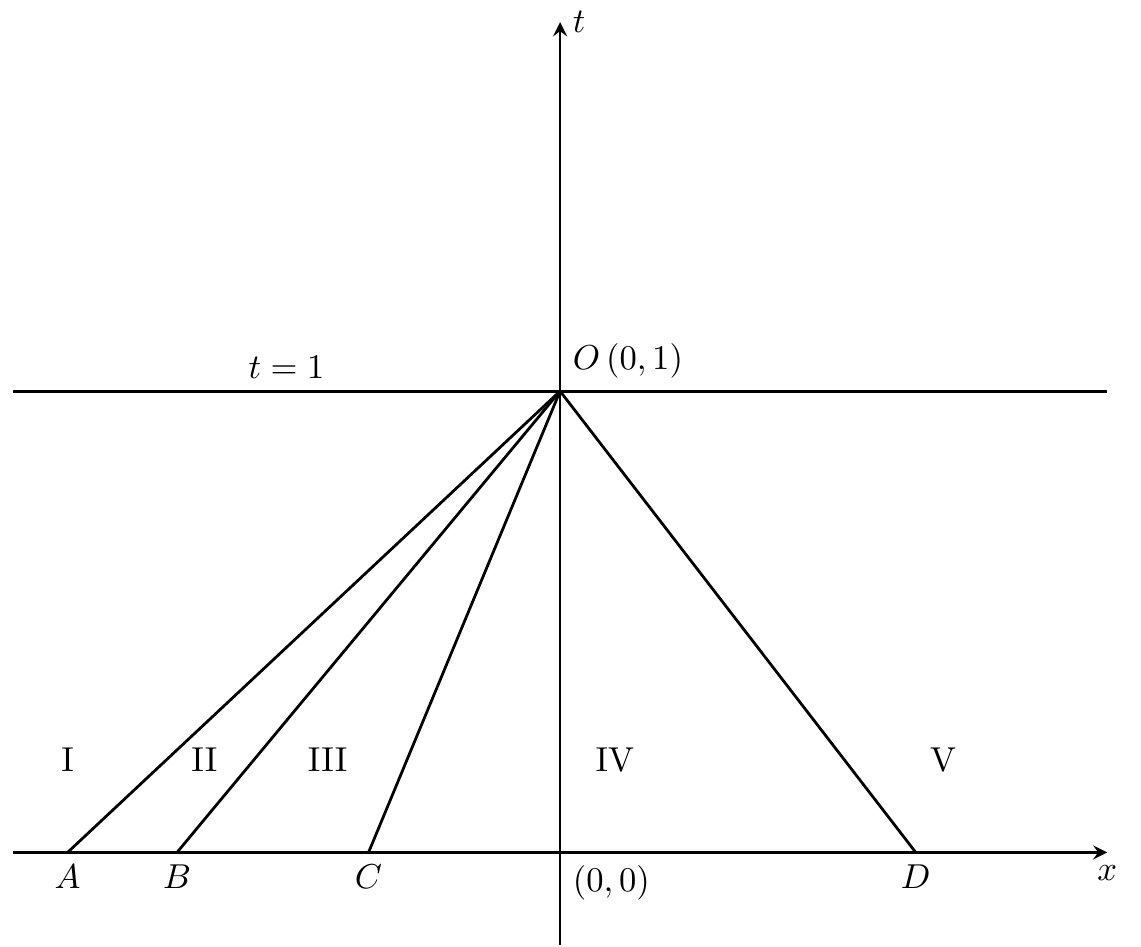}
    \caption{Cumulation of two self-focusing Riemann waves separated by a constant flow in zone III, $\rho_3>0, u_3>0$.}
    \label{fig:fig4}
\end{figure}

Assuming that inequality \ref{eqn:ineq3} is not satisfied and, therefore, the value of speed is bounded from above by the corresponding expression, let us present the values of physical quantities in each of the five regions obtained (Fig. \ref{fig:fig4}).

In region I, we have a constant flow and the following values of the variables $u=u_1,p=p_1,c=c_1$.
In region II there is a self-focusing Riemann R-wave and, therefore, the following relations hold:
\begin{equation*}
    R \equiv u-\frac{2}{\gamma-1}c=\text{const}
\end{equation*}
\begin{equation*}
    u_{II}-\frac{2}{\gamma-1}c_{II}=u_{1}-\frac{2}{\gamma-1}c_{1}
\end{equation*}
Region III is a constant flow and, therefore, $p=p_{III} \equiv \text{const}, c=c_{III} \equiv \text{const}, u=u_{III} \equiv \text{const}$. In region IV there is a self-focusing Riemann S-wave and, therefore, the following relations hold:
\begin{equation*}
    S \equiv u+\frac{2}{\gamma-1}c=\text{const}
\end{equation*}
\begin{equation*}
    u_{IV}+\frac{2}{\gamma-1}c_{IV}=u_0+\frac{2}{\gamma-1}c_0=\frac{2}{\gamma-1}c_0
\end{equation*}
Region V is a constant flow: $u=u_0=0, p=p_0, c=c_0$.
Equations defining rays $OA, OB, OC, OD$ have the form:
\begin{equation*}
    \frac{dx}{dt}=u_1+c_1,\quad \frac{dx}{dt}=u_{III}+c_{III},\quad \frac{dx}{dt}=u_{III}-c_{III},\quad \frac{dx}{dt}=c_0
\end{equation*}
Let us write the conditions for this configuration compatibility: $u_1+c_1>u_{III}+c_{III}>u_{III}-c_{III}>c_0$.
The compatibility conditions are trivial in the case of a single set of constants $\left(u_{III}, c_{III}\right), u_{III}>0, c_{III}<c_0$, satisfying the constancy conditions of the corresponding Riemann invariants:
\begin{equation*}
    c_{III}=\frac{1}{2}\left( c_1+c_0\right) - \frac{\gamma-1}{4}u_1
\end{equation*}
\begin{equation*}
    u_{III}=\frac{1}{\gamma-1}\left( c_0-c_1\right) + \frac{1}{2}u_1
\end{equation*}
\begin{equation}   \label{eqn:ineq4}
    u_1 > \frac{2}{\gamma - 1} c_1 \left ( 1 - {p'}^\frac{1-\gamma}{2 \gamma} \right )
\end{equation}    
Thus, at speeds satisfying condition \ref{eqn:ineq4} and not satisfying condition \ref{eqn:ineq3}, all step discontinuities are formed isentropically as a result of merging of two self-focusing Riemann waves in a simply connected polytropic medium. \\
The above calculations allow us to formulate three important statements. \\    
\textbf{Lemma 1.} The velocity jump at an arbitrary discontinuity formed in a polytropic environment by an isentropic process is bounded from below and must exceed:
\begin{equation*}
    \Delta u_{cr} \equiv \frac{2}{(\gamma-1)} c_1 \left[ 1 - {p'}^{\frac{(1-\gamma)}{2 \gamma}} \right]
\end{equation*}
Therefore, the Mach number $M_1 \equiv \frac{u_1}{c_1}$ of the incident flow is also limited:
\begin{equation*}
    M_1 \geq 2 \frac{1 - {p'}^{\frac{1-\gamma}{2 \gamma}}}{\gamma -1}
\end{equation*}
\textbf{Lemma 2.} An arbitrary discontinuity, formed by an isentropic process, and carrying a speed jump exceeding
\begin{equation*}
    \Delta u_{cr} \equiv \frac{2}{\gamma-1} c_1 \left[ 1 + {p'}^{\frac{1-\gamma}{2 \gamma}} \right]
\end{equation*}
is a consequence of the violation of simply connectedness that forms its polytropic medium. \\
\textbf{Lemma 3.} A jump can be formed by a non-isentropic process only in a contracting polytropic medium: $M_1>0$

The statement contained in Lemma 3 is a direct consequence of the prohibition of expansion shock waves, which in turn follows from the second law of thermodynamics, postulating an increase in entropy of a conservative thermodynamic system. \par
The paper discusses density jumps of limited intensity, because otherwise a jump, arising and disappearing at the same moment in time, is degenerate. Indeed, within the framework of the accepted conditions, the unboundedness of the density jump means the absence of a medium in the semi-infinite region to the right, that is, the density is zero to the right of the discontinuity. Such a jump is realized in the framework of self-similarity in a unique way: the right self-focusing Riemann wave propagating into the vacuum region. After focusing such a wave at the point of formation of the jump, the newly formed jump instantaneously evolves into a centered expansion wave, which is turned to the left and borders with the vacuum zone. This degenerate jump is of independent interest since its existence goes against the generally accepted point of view that the intersection of characteristics of one family necessarily leads to the production of entropy and, therefore, to a discontinuity of a flow. \par
The formulated statements exhaust the existence conditions of self-similar perturbations that form jumps in polytropic media.

\section{Formation of a solitary shock wave by two self-focusing Riemann waves in a simply connected region.}
We consider the solitary shock wave as a result of the evolution of an arbitrary jump and find the conditions under which the solitary shock wave is realized. We use the solution of an arbitrary discontinuity decay problem. The solution to this problem contains the following configurations: two expansion waves; two shock waves separated by a contact discontinuity; an expansion wave and a shock wave; a shock wave and an expansion wave; as well as the conditions for the implementation of each configuration. A solitary shock wave is the boundary of two configurations: two shock waves separated by a contact discontinuity and an expansion wave-shock wave separated by a contact discontinuity. The intersection of the conditions for the realization of these two configurations gives the condition for the appearance of a solitary shock wave. Since the shock wave can be attributed to both configurations under the condition that the left shock wave degenerates into a constant flow in a configuration of two shock waves, and the expansion wave degenerates into a constant flow in the configuration of an expansion wave and a shock wave. \par
In this section, we obtain the conditions for the existence of self-similar perturbations that form a solitary shock wave as a result of an isentropic process, both in a simply connected region and in the region containing the vacuum zone. The conditions obtained will allow us to introduce a natural classification of solitary shock waves by origin. \par
The constraints on the flow velocity obtained in lemmas 1 and 2 are satisfied for any jumps of physical quantities. Consequently, taking into account the relation 1, which distinguishes solitary shock waves from the general class of jumps, these restrictions will ensure the realization of a given perturbation configuration. \par
Let us rewrite the expression for the speed, defined by the equation \ref{eqn:ineq1}, in the form:
\begin{equation*} 
    u_1=(1-\varkappa)c_0 \left[ \sqrt{\frac{p'+\varkappa}{1+\varkappa}}-\sqrt{\frac{1+\varkappa}{p'+\varkappa}} \right]
\end{equation*}
under conditions \ref{eqn:ineq3} and \ref{eqn:ineq4}:
\begin{equation*} 
    \frac{2}{\gamma-1}c_1 \left[ 1- p'^{\frac{1-\gamma}{2 \gamma}} \right] \leq u_1 \leq \frac{\gamma-1}{2 \gamma}\left(c_1+c_0 \right)
\end{equation*}
Then performing a chain of calculation we get:
\begin{equation*} 
    \frac{2(1-\varkappa)}{\gamma-1} \left[{p'}^{\frac{\gamma-1}{2 \gamma}} -1 \right] \leq \frac{p'-1}{\sqrt{(p'+\varkappa)(1+\varkappa)}} \leq 
\end{equation*}
\begin{equation*} 
    \leq \frac{(\gamma -1)(1-\varkappa)}{2 \gamma} \left( 1+ {p'}^{\frac{\gamma-1}{2 \gamma}}\right)
\end{equation*}
In view of
\begin{equation*} 
    \gamma=\frac{k+2}{k}, \quad \frac{\gamma-1}{2 \gamma} = \frac{1}{k+2}
\end{equation*}
and
\begin{equation*} 
    \varkappa=\frac{\gamma-1}{\gamma+1}=\frac{1}{k+1}, \quad 1+\varkappa= \frac{k+2}{k+1},\quad 1-\varkappa = \frac{k}{k+1},
\end{equation*}
where $k$ is a rheological parameter that determines the number of degrees of freedom of the molecules that make up the continuous medium, we finally get \ref{eqn:ineq5}.
\begin{figure*}[bt!]
\begin{equation} \label{eqn:ineq5}
    \sqrt{\frac{k+2}{(k+1)^3}}k^2 \left [ {p'}^{\frac{1}{k+2}} - 1 \right ] \leq \frac{p'-1}{\sqrt{p'+\varkappa}} \leq \sqrt{\frac{1}{(k+3)^3 (k+2)}} k \left ({p'}^{\frac{1}{k+2}}+1 \right )
\end{equation}
\end{figure*}
The first part of expression \ref{eqn:ineq5} allows us to formulate the following statement. \\
\textbf{Lemma 4.} Any solitary shock wave appears in the isentropic flow of a polytropic medium as a result of the cumulation of two pseudo-centered (self-focusing) Riemann waves PCR, PCS.
The proof follows from the fulfillment of the first part of inequality \ref{eqn:ineq5}:
\begin{equation*}
\sqrt{\frac{k+2}{(k+1)^3}}k^2 \left [ {p'}^{\frac{1}{k+2}} - 1 \right ] \leq \frac{p'-1}{\sqrt{p'+\varkappa}}
\end{equation*}
for any $p' \geq 1$. Indeed $p'=1$ is root of equation 
\begin{equation*}
\sqrt{\frac{k+2}{(k+1)^3}}k^2 = \frac{p'-1}{\left [ {p'}^{\frac{1}{k+2}} - 1 \right ] \sqrt{p'+\varkappa}}
\end{equation*}
and function  
\begin{equation*}
F(p') \equiv \frac{p'-1}{\left [ {p'}^{\frac{1}{k+2}} - 1 \right ] \sqrt{(p'+\varkappa)}}
\end{equation*}
monotonically increases. \\
It follows from the second part of inequality \ref{eqn:ineq5} that in the case of its violation the simple connectedness of the region breaks and the following statement holds: \\
\textbf{Lemma 5.} Any solitary shock wave with sufficient intensity $p' \geq p_{cr}'$ appears in the isentropic flow of a polytropic medium as a result of the cumulation of two pseudo-centered Riemann waves PCR, PCS separated by a vacuum region. 
\begin{equation*}
\frac{p'-1}{\sqrt{{\rho}'+\varkappa}} \geq \sqrt{\frac{1}{(k+1)^3 (k+2)}} k \left ({p'}^{\frac{1}{k+2}}+1 \right )
\end{equation*}
is true $\forall p' \geq p_{cr}' > 1$, where $p_{cr}'$ is the only root of equation 
\begin{equation*}
f(p') \equiv  \frac{p'-1}{\left ({p'}^{\frac{1}{k+2}}+1 \right ) \sqrt{p' + \varkappa}}=\sqrt{\frac{1}{(k+1)^3 (k+2)}}k
\end{equation*}
with monotonically increasing function $f(p')$, $p'>1$ on the left side.
It is shown that any solitary shock wave can be formed by an isentropic flow and in a unique way within the framework of self-similarity. The perturbation configurations obtained above, which specify the flow data and their realization conditions allow us to distinguish two classes of shock waves as physical phenomena. We call the shock waves generated by disturbances containing a vacuum zone — high-entropy shock waves, since the entropy jump in those is bounded only from below. The shock waves arising due to the evolution of perturbation occupying a simply connected region of space are simply connected shock waves. The entropy jump in this class varies starting from infinitely small, and is bounded from above. This classification is the result of their occurrence mechanism consideration. \par
\begin{figure}
    \centering
    \includegraphics[width=.7\linewidth]{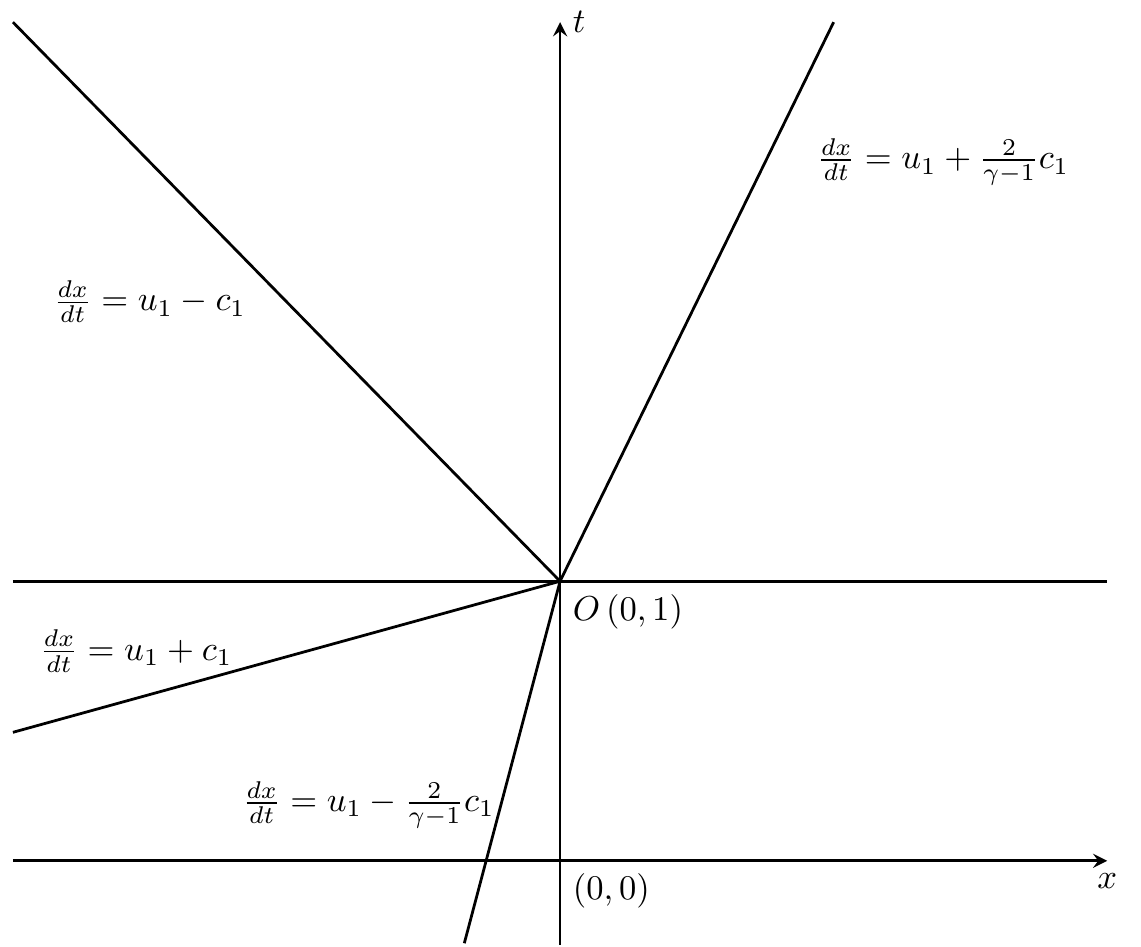}
    \caption{Evolution of the right self-focusing Riemann wave to the left centered Riemann wave, without the formation of a shock front. Region III is vacuum: $\rho_3=c_3 \equiv 0, u_+=\frac{\gamma-1}{2}u_1-c_1, u_-=\frac{\gamma-1}{2}u_1-c_1$.}
    \label{fig:fig5}
\end{figure}

Thus, under the assumption of the flow self-similarity there is a single mechanism for the occurrence of a solitary shock wave, realized by the process of accumulation of self-focused Riemann waves. Therefore, a solitary shock wave cannot arise as a result of the occurrence of only one Riemann wave. Moreover, not every occurrence of the Riemann wave leads to the appearance of a flow containing a shock front. Fig. \ref{fig:fig5} shows an example of a right self-focusing Riemann wave evolving into an isentropic current containing only a left expansion wave.

\section{Parametric dependence of a self-similar flow wave pattern on the velocity of the incident flow.}
The results obtained allow us to describe the dynamics of a wave pattern of the self-similar flow that forms a jump, as a function of the velocity of the incident flow. Indeed, for each specific pair of density values at a jump, it is possible to consider a semi-infinite range of the incident flow velocity variation as a parameter controlling the type of wave pattern. Negative values of speed are excluded from consideration, since as shown above the formation of jumps is impossible in a stretching medium in principle, even in non-isentropic processes. So, for  velocities in the interval $(0, u_\alpha)$, he wave pattern corresponds only to a non-isentropic process and is a sequence of self-focusing Riemann waves and shock fronts separated by a contact discontinuity. Thus, all left shock fronts and Riemann waves are to the left of it and all the right ones are to the right. It can be shown that for any flow that is self-similar, constant at infinity and contains a perturbation region consisting only of a self-focusing Riemann wave and a contact discontinuity, it is possible to offer a countable set of self-similar flows with identical values at infinity and perturbations inside, consisting of a set of shock fronts, self-focusing Riemann waves and a contact discontinuity occupying a strictly smaller region than in the isentropic process. Hence, in a non-isentropic process, even under the condition of self-similarity, the perturbation connecting two constant flows is not unique. Evidently, non-isentropic processes allow for the construction of self-similar perturbations for the entire positive semi-infinite interval of velocity variations in a non-unique way. However, the condition of constant entropy in the entire region of a jump formation ensures the uniqueness of a possible perturbation in the framework of self-similarity. \par
If velocity of the incident flow lies within the interval $(u_\alpha, u_\beta)$, the wave picture of isentropic process consists of two self-focusing Riemann waves. For a velocity value exactly falling on the left boundary of the interval, the left Riemann wave (on the right) degenerates into a constant flow. With an increase in the velocity of the incident flow, the fronts of self-focused Riemann waves approach each other, until velocity of the incident flow becomes exactly the right boundary of the interval. At this point, the value of density and speed of sound on the merged fronts becomes zero. A further increase in the flow velocity in the interval $(u_\beta, + \infty)$ leads to the expansion of the vacuum region due to reflection of the front of right self-focusing wave, while the front of left self-focusing wave separating it from the vacuum zone loses sensitivity to the parameters of the incident flow and remains fixed for any velocity from the indicated semi-infinite interval.

\section{Conclusion}
In this work we studied the mechanisms of isentropic processes of jumps formation in polytropic media. The properties of finite continuous self-similar solutions, self-focusing Riemann waves, are studied. The problem of a gas-dynamic jump formation in a polytropic medium is stated and solved. Wave patterns of perturbations forming jumps are found and the conditions for their realization are obtained. The critical values of incident flow velocity that divide the range of parameters into intervals corresponding to each wave pattern are presented. It is shown that there is a critical velocity, below which a discontinuity cannot be formed isentropically. Also found is the second critical value of velocity, exceeding which a discontinuity is formed only in the presence of a vacuum region. It is shown that any solitary shock wave arises in the isentropic flow as a result of the occurrence of two self-focusing Riemann waves. The conditions are found under which the region of solitary shock wave formation contains a vacuum region. The division of shock waves into two classes, high-entropy and simply connected, based on the mechanism of their occurrence is proposed. All the results discussed in the paper were obtained for the case of nonstationary one-dimensional gas dynamics; however, it should be noted that these results are generalized in a natural way to the case of an arbitrary quasilinear system of hyperbolic partial differential equation which describe for example, the two-dimensional stationary flows of compressible gas. As another example, we can write the value of the critical speed separating the wave patterns of hydraulic jump formation described by the shallow water equations \cite{batchelor1967introduction}.

The work was supported by the Russian Foundation for Basic Research (project no. 19-02-00016).

\bibliographystyle{unsrt}  
\bibliography{references}  

\end{document}